\documentclass[12pt]{article}
\usepackage{graphicx}
\usepackage{amsmath}
\usepackage{longtable}

\begin{document}

\makeatletter
\renewcommand*{\@cite}[2]{{#2}}
\renewcommand*{\@biblabel}[1]{#1.\hfill}
\makeatother

\title{Variations of the Interstellar Extinction Law within the Nearest Kiloparsec}
\author{G.~A.~Gontcharov\thanks{E-mail: georgegontcharov@yahoo.com}}

\maketitle

Pulkovo Astronomical Observatory, Russian Academy of Sciences, Pul\-kov\-skoe sh. 65, St. Petersburg, 196140 Russia


Multicolor photometry from the Tycho-2 and 2MASS catalogues for 11 990 OB and 30 671 K-type
red giant branch stars is used to detect systematic large-scale variations of the interstellar extinction
law within the nearest kiloparsec. The characteristic of the extinction law, the total-to-selective extinction
ratio $R_V$, which also characterizes the size and other properties of interstellar dust grains, has been
calculated for various regions of space by the extinction law extrapolation method. The results for the two
classes of stars agree: the standard deviation of the ``red giants minus OB'' $R_V$ differences within 500 pc of
the Sun is 0.2. The detected $R_V$ variations between 2.2 and 4.4 not only manifest themselves in individual
clouds but also span the entire space near the Sun, following Galactic structures. In the Local Bubble
within about 100 pc of the Sun, $R_V$ has a minimum. In the inner part of the Gould Belt and at high Galactic
latitudes, at a distance of about 150 pc from the Sun, $R_V$ reaches a maximum and then decreases to its
minimum in the outer part of the Belt and other directions at a distance of about 500 pc from the Sun,
returning to its mean values far from the Sun. The detected maximum of $R_V$ at high Galactic latitudes is
important when allowance is made for the interstellar extinction toward extragalactic objects. In addition,
a monotonic increase in $R_V$ by 0.3 per kpc toward the Galactic center has been found near the Galactic
equator. It is consistent with the result obtained by Zasowski et al. (2009) for much of the Galaxy. Ignoring
the $R_V$ variations and traditionally using a single value for the entire space must lead to systematic errors
in the calculated distances reaching 10\%.

\newpage
\section*{THE METHOD}

The interstellar extinction law describes the dependence
of extinction $A_{\lambda}$ on emission wavelength $\lambda$.
Since the extinction is difficult to measure directly at
various wavelengths, the reddening of a star, i.e., the
deviation of its color from the true one due to selective
extinction, is traditionally measured to determine
the extinction law. In the visual spectral range, the
reddening $E_{(B-V)}$ is commonly considered. The
reddening is currently believed to be produced by
dust grains whose size is smaller than the emission
wavelength. In the visual spectral range, these grains
are less than 1 micron in size.

The extinction, for example, in the $V$ band is
determined by taking into account the reddening:
$A_V=R_V\cdot E_{(B-V)}$. The coefficient $R_V$ is the total-to-selective extinction ratio. The same (nonselective
or ``gray'') extinction at all wavelengths is apparently
produced by dust grains with a size larger than the
wavelength, i.e., more than 1 microns, for the visual spectral
range. Therefore, $R_V$ and similar coefficients for
other wavelengths (for example, $A_{Ks}/E_{(H-Ks)}$),
first, reflect the fraction of coarse dust in the absorbing
matter and the mean dust grain size and, second,
possibly other dust grain properties. Since these
coefficients, the wavelength, and the grain size are
related between themselves, any of these coefficients
is apparently a single versatile characteristic of the
extinction law in a particular region of space and
time (Fitzpatrick and Massa 2007). Given that the
extinction law is often determined at present not for
the visual range but for the infrared in the form of a coefficient
similar to $R_V$, below we use the more general
name ``extinction law'' instead of the ``coefficient $R_V$''.

The reddening of a star is traditionally measured
as the difference of the observed color and the zero
point -- the color of an unreddened star of the same
spectral type. As was shown in the review by Perryman
(2009), the natural scatter of colors for unreddened
stars of the same spectral subtype (even if we
forget the common classification errors) is typically
$0.3^m$. Because of such a low accuracy, this method
becomes a thing of the past in the cases where the
individual reddenings or extinctions are averaged for
a group of stars.

\subsection*{Using Present-Day Surveys}

The new, basically statistical method based on the
all-sky photometric surveys of millions of stars that
have appeared in recent years gives a much smaller
error. These surveys include the Tycho-2 catalogue
(H\o g et al. 2000) containing photometry in the $B_T$
and $V_T$ bands with the effective wavelengths $\lambda_{B_T}=0.435$ and $\lambda_{V_T}=0.505$ microns
(for comparison, Johnson's
$V$ band has $\lambda_{V}=0.553$ microns), the 2MASS catalogue
(Skrutskie et al. 2006) with infrared photometry
in the $J$ ($\lambda_{J}=1.24$ microns), $H$ ($\lambda_{H}=1.66$ microns), $Ks$ ($\lambda_{Ks}=2.16$ microns)
bands (for comparison, Johnson's
$K$ band has $\lambda_{V}=2.22$ microns), and other surveys.
Owing to accurate photometry, the stars from such
surveys are grouped by classes on color--color and
color--magnitude diagrams. This allows the stars of
certain classes to be selected and the mean characteristics
of the selected stars in sky fields or spatial cells
can be calculated with confidence owing to the large
number of stars, despite the presence of admixtures.
The mean color for stars of a certain class in each cell
is among the characteristics being determined. The
mean color of stars close to the Sun may be considered
unreddened and can be used as a zero point: the
difference of the colors in a spatial cell far from the
Sun and in the solar neighborhood is the mean stellar
reddening in that cell. Applying this method in various
modifications allowed red giant branch (RGB)
stars (Dutra et al. 2003), red giant clump (RGC) stars
(Drimmel et al. 2003; Gontcharov 2008b; Zasowski
et al. 2009), OB stars (Gontcharov 2008a), and
F-type dwarfs and subgiants (Gontcharov 2010) to
be selected. As a result, Dutra et al. (2003) constructed
a three-dimensional reddening map for stars
toward the Galactic center, Drimmel et al. (2003)
constructed a three-dimensional reddening map for
stars at heliocentric distances $2<r<8$ kpc, and
Gontcharov (2010) constructed a $E_{(J-Ks)}$ reddening
map with an accuracy of $0.03^m$ for stars within
1500 pc of the Sun.

One would think that the same multicolor photometry
from large-scale surveys can be used to determine
the extinction law using color--color diagrams.
Indeed, the ratio of two colors for a star is
a characteristic of its spectral energy distribution (a
set of ratios of various colors in distinctly different
spectral regions is better). For a group of stars with
approximately the same spectral energy distribution
(stars of the same class) but with different reddenings,
reflects the change in the spectral energy distribution
with reddening. A large change in the energy distribution
(large extinction) at a small reddening corresponds
to high values of $R_V$; a small change in the
energy distribution at a large reddening corresponds
to low values of $R_V$.

In other words, given that the infrared extinction
is considerably lower than the visual one or, more
specifically, $A_{K}\approx0.1A_V$, we have
$R_V=A_V/E_{(B-V)}=1.1(A_V-A_K)/E_{(B-V)}=1.1E_{(V-K)}/E_{(B-V)}$ (Straizys 1977).

Given that, for example, $E_{(B_T-V_T)}=A_{B_T}-A_{V_T}$, estimating $A_K\approx0.1A_V$, just as estimating
$R_V$, as a function of $E_{(V_T-Ks)}/E_{(B_T-V_T)}$ directly
or indirectly requires knowing the extinction
law, i.e., the wavelength dependence of extinction.
The numerous attempts to establish a
universal extinction law considered, for example,
by Fitzpatrick and Massa (2009) have led to the
detection of spatial variations in the law and even
questioned whether the power-law wavelength dependence
of extinction is valid. However, as was
noted by Zasowski et al. (2009), the mean ratios
of extinctions in different spectral ranges have so
far been determined with a relatively low accuracy.
For example, values from 0.09 to 0.11 for the ratio
$A_K/A_V$ and very different formulas for the dependence
of $R_V$ on $E_{(V_T-Ks)}/E_{(B_T-V_T)}$ are encountered
in the literature:
$R_V=1.12E_{(V-K)}/E_{(B-V)}+0.02$ (Fitzpatrick 1999),
$R_V=1.039E_{(V_T-Ks)}/E_{(B_T-V_T)}+0.02$ (Sudzius and Raudeliunas 2003),
$R_V=1.19E_{(V_T-Ks)}/E_{(B_T-V_T)}-0.26$ (Fitzpatrick and Massa 2007),
$R_V=1.36E_{(V_T-Ks)}/E_{(B_T-V_T)}-0.79$ (Fitzpatrick and Massa 2009).

The problem is that the mean ratios of extinctions
at different wavelengths are to a greater or lesser extent
empirical and, therefore, depend on the sample of
stars. Some uncertainty in the ``absolute'' calibration
of the derived $R_V$ will also remain in the results of this
study, but its variations found are not only real but
also exceed this uncertainty by several times.

A desirable approach to determining the extinction
law in future is to reconstruct the entire wavelength
dependence of extinction using homogeneous multiband
photometry (spectrophotometry is better) for
millions of stars with an accuracy of at least $0.01^m$
in the ultraviolet, visible, and infrared ranges in many
spatial cells of the Galaxy. However, the variations of
the extinction law can and should be revealed at the
existing accuracy level.

Having examined the $B$, $V$, $K$ (Moro and Munari 2000), $B_T$ , $V_T$ (H\o g et al. 2000), and $Ks$
(Skrutskie et al. 2006) filter profiles in combination with the typical profiles of the B5V-B7V and K3III
stars considered here (Pickles 1998), we find that the
ratio $E_{(V_T-Ks)}/E_{(B_T-V_T)}$ differs from $E_{(V-K)}/E_{(B-V)}$ mainly through a slight blueshift of
$V_T$ relative to $V$. Taking into account the typical
influence of a small reddening on the spectral ranges
under consideration as a set of monochromatic bands,
we find that the ratio $E_{(V_T-Ks)}/E_{(B_T-V_T)}$ differs from $E_{(V-K)}/E_{(B-V)}$ only slightly and
we can adopt
\begin{equation}
\label{evkebv}
R_V=1.12\cdot E_{(V_T-Ks)}/E_{(B_T-V_T)}.
\end{equation}
The coefficient was determined with a low accuracy
of $1.12\pm0.05$ due to the uncertainty in the extinction
law mentioned above. As has already been noted, this
also introduces an uncertainty in the final result, but
it is much smaller than the amplitude of the variations
found.

Proportionality (1) is also valid for the color differences
for a pair of stars with the same spectrum:
\begin{eqnarray*}
R_V&\sim(E_{(V_T-Ks)_2}-E_{(V_T-Ks)_1})/(E_{(B_T-V_T)_2}-E_{(B_T-V_T)_1})=\\
          &=((V_T-Ks)_2-(V_T-Ks)_0)-((V_T-Ks)_1-(V_T-Ks)_0)/\\
          &/((B_T-V_T)_2-(B_T-V_T)_0)-((B_T-V_T)_1-(B_T-V_T)_0)=\\
          &=((V_T-Ks)_2-(V_T-Ks)_1)/((B_T-V_T)_2-(B_T-V_T)_1).
\end{eqnarray*}
Thus, the reddening difference between two stars
with the same spectrum is equal to the difference of
their colors. Therefore, the coefficient of the linear
dependence of $(V_T-Ks)$ on $(B_T-V_T)$ for groups of
stars of the same class is $R_V/1.12$.

An example of the selection of stars of one class
with similar spectra was given previously (Gontcharov
2011): the dereddened colors of 30 671 selected
K-type RGB stars have a small scatter: $\sigma(V_T-Ks)_0=0.35^m$, $\sigma(B_T-V_T)_0=0.08^m$. An example
of determining the extinction law using these stars
is presented in Fig. 1. The $(V_T-Ks)$ -- $(B_T-V_T)$
diagram is shown for K-type RGB stars in the spatial
cell with coordinates $0^{\circ}<l<90^{\circ}$, $-10^{\circ}<b<+10^{\circ}$, $200<r_{HIP}<300$ pc, where $l$ is the
Galactic longitude, $b$ is the Galactic latitude, and $r_{HIP}$
is the distance from Hipparcos parallaxes (ESA 1997;
van Leeuwen 2007). The vertical and horizontal
bars mark the formal accuracy of the photometry.
The trend $(V_T-Ks)=2.76(B_T-V_T)-1.17$ was
drawn by the least-squares method. It may be concluded
that $R_V=1.12\times2.76=3.09$ in this spatial
cell, which corresponds to the widespread ``mean'' or
``standard'' estimate.

Straizys (1977, pp. 39--40) called this the extinction
law extrapolation method; Zasowski et al. 2009)
called this the color ratio method. This method was
apparently first applied by Jonhson and Borgman
(1963). Unfortunately, as Wegner (2003) showed,
this is the only direct method of determining $R_V$ using
individual stars rather than, say, clusters applicable at
present.

The difficulties in using the extinction law extrapolation
method are the following.
\begin{itemize}
\item 
As has been shown above, a sample of stars
with a similar spectral energy distribution, i.e.,
stars of the same spectral type, luminosity
class, and metallicity, is needed to reduce
the scatter of dereddened stellar colors. In
addition, the peculiar stars and stellar pairs
in which the photometry and spectra cannot
be considered separately for each component
should be excluded.
\item 
The larger the range of stellar colors in the spatial
cell under consideration, the more accurate
the result (although the minimal scatter of
dereddened colors is desirable!). Therefore, the
method is applicable only in sufficiently large
spatial cells, where the stars exhibit a wide
range of colors as a result of their reddening
and, in addition, the maximum reddening in
such a cell is great. To be more precise, the
method is applicable in a cell where the mean
reddening is larger than the natural scatter of
colors for unreddened stars and, in addition,
the errors in the colors are larger due to the
errors in the original photometry. For the stars
used here, this is the region of space with $A_V>0.25^m$. Consequently, the method does not
work within about 50 pc of the Sun. At high
Galactic latitudes, as we show below in the
results for RGB stars, the layer of absorbing
matter above and below the Sun provides the
necessary reddening and extinction $A_V\approx0.3^m$.
\item 
It follows from the aforesaid that the original
photometry for the stars used must be fairly
accurate. In any case, the colors must be
determined with an accuracy that is a factor
of $R_V$ higher than the mean level of $A_V$ in the
spatial cell being investigated. The accuracy of
the colors is at least $0.1^m$ -- a level at which
the method is efficient, while much more reliable
results can be obtained at a photometric
accuracy of $0.01^m$.
\item 
The stars used must be sufficiently numerous.
This is important not only for revealing and excluding
peculiar stars but also because a large
number of stars can to some extent compensate
for the inaccuracy of the photometry. For
example, nine stars in each spatial cell under
consideration are needed to calculate $R_V\approx3$
in the spatial cell with an accuracy of the original
photometry at least $0.1^m$, i.e., with a relative
accuracy of 3\%. Consequently, 9000 uniformly
distributed stars are needed to analyze $R_V$ in
the sphere with a radius of 500 pc at a cell
radius of 50 pc.
\item 
To investigate the regions with large extinction,
we must use high-luminosity stars seen
from far away.
\item 
To reduce the possible biases of the quantities
being determined, the stars must be distributed
fairly uniformly in space and the sample must
be to a large extent complete to some distance.
\end{itemize}

Even the first application of the extinction law extrapolation
method by Jonhson and Borgman (1963)
allowed not only large deviations of $R_V$ from 3.1 for
some stars but also a smaller (in amplitude) smooth
dependence of $R_V$ on Galactic longitude with a minimum
at $l\approx110^{\circ}$ to be detected. However, as a result
of the listed stringent requirements for the method,
researchers adopt a certain constant (in the entire
space) extinction law (for example, $R_V=3.1$), i.e.,
they assume the interstellar medium to be homogeneous
with regard to the dust grain properties, despite
progress in studying the reddening of stars. The
maps of reddening variations obtained in this case are
called ``extinction maps''. However, when comparing
such maps constructed from photometry at distinctly
different wavelengths, there are discrepancies whose
value apparently correlates with the dust temperature
(Dutra et al. 2003; Peek and Graves 2010). The
discrepancies between the reddening maps do not
allow the spatial variations of the extinction law to be
ignored any longer.

\subsection*{Previous Implementation of the Method}

Various researchers have found that $R_V$ within the
kiloparsec nearest to the Galactic center is considerably
smaller than that near the Sun (Popowski 2000).
However, only Sumi (2004) showed that this is not a
local anomaly of the extinction law but possibly largescale
variations within at least 1 kpc of the Galactic
center: when recalculated, $R_V$ changes here along $l$
but not along $b$ approximately by 0.2 kpc$^{-1}$ if the
stars being investigated are assumed to belong to a
bar oriented at an angle of about 45$^{\circ}$ to the Sun. In
this case, the near part of the bar in the first Galactic
quadrant shows a lower value of $R_V$ than its far part
in the fourth quadrant and there is no extremum expected
at the Galactic center. The grow of $R_V$ with
increasing heliocentric distance toward the Galactic
center is consistent with the results considered below,
but extrapolating the results from Sumi (2004) to
the circumsolar space gives too low values of $R_V$
and, in addition, the absence of an extremum at the
center defies common sense. Thus, the results by
Sumi (2004) require confirmation and analysis.

The large-scale spatial variations of the extinction
law over at least 12 kpc were established more reliably
by the extinction law extrapolation method and were
analyzed in detail by Zasowski et al. (2009) based
on a combination of 2MASS (three near-infrared
bands) and Spitzer-IRAC (four mid-infrared bands)
photometry for RGC stars. The seven photometric
bands used cover the wavelength range from 1.2 to
8 microns. In this case, RGC stars near the Galactic plane
in the bulk of the first and fourth Galactic quadrants
farther than 10$^{\circ}$ for the Galactic center (a total of
290 sq. deg. on the sky) in the range of magnitudes
$11^m<J<15.5^m$.
5 were selected on the color--magnitude $(J-Ks)$ -- $J$ diagram. Given $M_J\approx-1^m$
for RGC stars, they were selected at a heliocentric
distance $2.5<r<15$ kpc, not including the region
in the immediate vicinity of the Galactic center.
The sky region under consideration was divided into
$2.5^{\circ}\times 2^{\circ}$ cells, each containing from 820 to 60 000
selected stars. The colors with respect to the $H$ band
like $(J-H)$, $(H-Ks)$, etc. were used for each star.
For all stars in the cell, the linear dependence of the
remaining colors with respect to $(H-Ks)$ was found
by the least-squares method. The coefficient of this
dependence characterizes the infrared extinction law
and is related to $R_V$.

Systematic variations of the extinction law that
gave, when recalculated, $R_V$ variations between 3.1
and 5.5 were found by Zasowski et al. (2009) in the
Galactic region under consideration. The authors
pointed out the inaccuracy of recalculating the result
to $R_V$, because all of the previous absolute calibrations
of the extinction law based on the medium's homogeneity
are inapplicable. However, this inaccuracy
is smaller than the variations found by several times.
Since the longitude in the Galactic region being investigated
strongly correlates with the Galactocentric
distance, Zasowski et al. (2009) treat the longitude
dependence of the extinction law as a dependence
on the Galactocentric distance and, consequently, as
a systematic decrease in the dust grain size and a
decrease in nonselective extinction with increasing
Galactocentric distance. Thus, dust of supermicron
and submicron sizes apparently dominates in the central
and outer Galactic regions, respectively. A correlation
of the dust grain size with the stellar metallicity,
the grain chemical composition and shape, and other
parameters that can depend on the Galactocentric
distance is also possible. Zasowski et al. (2009)
point out that even excluding the known regions with
anomalously large $R_V$ from consideration does not
remove the systematic trend found. Thus, the $R_V$
variations were detected in much of the Galaxy and
not only in one or more anomalous regions. This
means that the $R_V$ variations are actually inherent in a
diffuse medium, not in dense clouds, and these variations
should not be confused with the well-known deviations
of $R_V$ from 3.1 in small star-forming regions.

Applying a universal extinction law (for example,
$R_V=3.1$) can give large systematic and random errors
when calculating the extinctions, distances, absolute
magnitudes, and other characteristics of stars
in any Galactic region. These errors as a function of
the error in $R_V$ were given by Reis and Corradi (2008):
for variations within $R_V$ $\pm1.5$ of the mean, the calculated
distances and/or magnitudes of stars are in error
by 10\%. In this case, the systematic $R_V$ variations
produce the systematic errors in the distances.

Zasowski et al. (2009) point out that the method
of extrapolating the extinction law using only infrared
photometry (without any visual bands) is applicable
only in regions with a very large extinction:
$E_{(J-Ks)}>0.35$, or $A_V>2^m$. In addition,
it is inapplicable near the Sun due to the absence
of accurate infrared photometry for bright stars (see
Gontcharov 2011). Only a combination of infrared
and visual photometry is possible within 1.5 kpc of
the Sun.

Thus, the above peculiarities of the theory and the
realization of the extinction law extrapolation method
at present leave a very narrow choice of original material
for investigating the variations of the extinction
law within several hundred pc of the Sun: OB, RGC,
and type-K RGB stars as numerous high-luminosity
stars. The Tycho-2 catalog is the only source of
their accurate visual photometry over the entire sky,
while the 2MASS catalog serves as the only source
of infrared photometry for such bright stars, although
there are also other sources of photometry for analyzing
the variations of the extinction law not over the
entire sky but in individual regions.

\section*{ORIGINAL DATA}

To analyze the variations of the extinction law, we
use the $B_T$, $V_T$, and $Ks$ photometry as well as the
trigonometric, $r_{HIP}$, and photometric, $r_{ph}$, distances
separately for the samples of OB and type-K RGB
stars. Our main goal is to detect consistent $R_V$
variations for so different classes of stars.

\subsection*{OB Stars}

The sample of 37 485 OB stars was obtained previously
(Gontcharov 2008a). To investigate the extinction
in the Gould Belt, we excluded the stars
with photometry poorer than $0.05^m$ at least in one of
the bands used from the sample (Gontcharov 2009).
In addition, the stars with an extinction exceeding
$A_V=1.1/(0.25+|\sin(b)|)$, as a rule, late-type peculiar
stars, were rejected. To increase the accuracy,
we excluded the stars with known spectral classification
from the Tycho Spectral Types catalogue (TST,
Wright et al. 2003) that did not belong to the O,
B, and A0 types from the sample. As a result (see
Gontcharov 2009), the sample of 15 670 remaining
OB stars was used. However, the spectral energy distribution
for A0 stars differs noticeably from that for
O and B ones. Therefore, for our study, we excluded
the A0 stars according to the TST from the sample;
11 990 OB stars remained in our sample.

The median accuracy of $(B_T-V_T)$ and $(V_T-Ks)$ is $0.03^m$. For all $\sigma(B_T-V_T)<0.07^m$, $\sigma(V_T-Ks)<0.06^m$.
Thus, the accuracy $\sigma(R_V)<0.1$ in 99\% of the spatial cells is achieved when only
six OB stars are present in the cell. However, based
on the available data, we cannot reliably exclude the
giants, supergiants, Be stars, and stars with a peculiar
spectrum, for which the reddening and extinction
can differer from the normal ones, from the sample.
To reduce the influence of these stars on the result,
the spatial cells were chosen to have at least 25 stars
in the cell. The high Galactic longitudes at which
there are almost no OB stars constitute an exception.
The method is also difficult to apply, because
the mean reddening of the OB stars under consideration
reaches an appreciable value, $\overline{E_{(B_T-V_T)}}>0.1$, only at $|b|<50^{\circ}$. Therefore, when using spherical
coordinates, the regions $b<-40^{\circ}$ and $b>+40^{\circ}$ were
considered as two large indivisible spatial cells. When
using rectangular $XYZ$ coordinates for the regions
$50<|Z|<150$ pc at $\sqrt{X^2+Y^2}<300$ pc, we interpolated
the results of the neighboring cells at the
same $Z$ and did not consider the cells with $|Z|>150$.
However, fairly reliable results were obtained for high
latitudes from RGB stars.

For our study, we performed a new calibration
of the absolute magnitude $M_{V_T}$ as a function of
$(B_T-V_T)_0$ based on 871 OB stars that, according
to Hipparcos, are closer than 300 pc. In this case,
$M_{V_T}=V_T+5-5\log(r_{HIP})-A_V$, where $A_V$ is the
extinction calculated previously (Gontcharov 2008a).
The $(B_T-V_T)_0$ colors were obtained by correcting
the original $(B_T-V_T)$ for the reddening that was
also calculated previously (Gontcharov 2008). Monte
Carlo simulations (Gontcharov 2011) showed that
the Malmquist and Lutz--Kelker biases may be neglected
for so close stars.

The distribution of these stars on the $(B_T-V_T)$ -- $M_{V_T}$ diagram is shown in Fig. 2: (a) with
the original $(B_T-V_T)$ and (b) with the dereddened
$(B_T-V_T)_0$. We see that the stars under consideration
are noticeably reddened. Dereddening
reduced the standard deviation from $\sigma(B_T-V_T)=0.13^m$ to $\sigma((B_T-V_T)_0)=0.08^m$.
and the mean from
$\overline{(B_T-V_T)}=0^m$ to $\overline{((B_T-V_T)_0}=-0.13^m$ (for all
11 990 stars, these quantities are
$\sigma(B_T-V_T)=0.2^m$, $\sigma((B_T-V_T)_0)=0.09^m$, $\overline{(B_T-V_T)}=0.15^m$, $\overline{((B_T-V_T)_0}=-0.09^m$).
These changes suggest that
the calculated reddening is plausible. In addition, as
has been noted above, the small $\sigma((B_T-V_T)_0)=0.08^m$, which does not exceed the mean reddening
$\overline{E_{(B_T-V_T)}}$ in the cells under consideration, is important for the method being applied.

The solid curve in Fig. 2 indicates the isochrone
for solar-metallicity stars with an age of 60 Myr typical
of late-OB stars calculated from the evolutionary
models of Girardi et al. (2000) by taking into account
the relations $(B_T-V_T)=(B-V)/0.85$ and $M_{V_T}=M_{V}+0.09(B_T-V_T)$ (ESA 1997). Given the errors
of the original photometry, this isochrone fits well
the bulk of the cloud of points after dereddening in
Fig. 2b. However, the larger number of points above
the isochrone suggests an admixture of giants in the
sample. The positions of the extremely blue stars
($(B_T-V_T)_0<-0.25$) are apparently explained by
photometric errors. Both giants and photometric
errors are inherent in the entire sample of 11 990
OB stars. They should be taken into account in
the calibration. Therefore, we adopted an empirical
calibration found by the least-squares method from
data for 871 stars under consideration. It is indicated
by the dashed line in Fig. 2b: $M_{V_T}=5.85(B_T-V_T)_0+0.84$. The accuracy of the original photometry
and parallaxes allows the accuracy of the coefficients
found to be estimated as 5.85$\pm$0.05 and 0.84$\pm$0.05,
respectively. The standard deviation for the stars
relative to this calibration straight line is $\sigma(M_{V_T})=0.8$. This allows the photometric distances
$\log(r_{ph})=(V_T-M_{V_T}+5-A_V)/5$ to be calculated with a relative
accuracy of 40\%.

The photometric distances are important in our
study, because the sample contains only 3606 Hipparcos
stars (30\%), the accuracy of $r_{HIP}$ is higher
than 40\% only for 1891 stars from them (16\%). Thus,
the sample of 11 990 under consideration is complete
or almost complete in a much larger region of
space (within 400 to 800 pc of the Sun, depending
on b) than the sample of Hipparcos stars (150 to
300 pc, respectively), while the photometric distances
for 84\% of the sample stars are more accurate than the
trigonometric ones. Therefore, below when considering
the variations of the extinction law, the results
using $r_{ph}$ are the main ones, while those with $r_{HIP}$ are
considered only for checking.

\subsection*{Type-K RGB Stars}

The sample of 30 671 K-type RGB stars was obtained
previously (Gontcharov 2011). The median
accuracy of $(B_T-V_T)$ and $(V_T-Ks)$ is 0.03$^m$ and 0.05$^m$, respectively. For 98\% and 84\% of the stars,
$\sigma(B_T-V_T)<0.1^m$ and $\sigma(V_T-Ks)<0.1^m$, respectively.

The absence of accurate near-infrared photometry
for bright stars ($Ks<5.5^m$) in modern astronomy is
a serious problem. For example, for the 2MASS
project, these stars turned out to be too bright and,
despite special efforts, even the formal accuracy of
their photometry is, on average, lower than 0.3$^m$; actually,
it can be even lower.

However, our cross-identification of the RGB
stars under consideration using the SIMBAD
database in Strasbourg showed that the IRAS satellite
(IRAS 1988) measured the infrared flux at wavelengths
of 12 and 25 microns for many of them. For stars
with fairly accurate measurements of the $Ks$ magnitude
and the 12-microns flux, these results were compared
and a correlation was found: $Ks=0.0087(\log_{2.512}(F12))^3)-0.0371(\log_{2.512}(F12))^2)-0.95(\log_{2.512}(F12)))+4.2$,
where $F12$ is the logarithm
of the infrared flux at 12 microns to base 2.512. The
scatter of $Ks$ magnitudes relative to this calibration
curve for stars with accurate $Ks$ and $F12$ photometry
is $\sigma(Ks)=0.18^m$. The calibration must give the same
accuracy for stars with inaccurate $Ks$ photometry.
Having applied this calibration to bright stars, we
retain them in the sample.

Thus, the accuracy $\sigma(R_V)<0.1$ is achieved everywhere
when 36 K-type RGB stars are present
in the cell. This condition can be fulfilled even for
high Galactic latitudes. The type-K RGB stars have
approximately the same $M_{V_T}$ as the OB stars considered.
Therefore, the sample of K-type RGB stars
is complete in the same space.

The photometric distances of the type-K RGB
stars under consideration were calculated previously
(Gontcharov 2011) with a relative error of 40\%. The
sample contains 9742 Hipparcos stars (32\%); the
accuracy of $r_{HIP}$ is higher than 40\% only for 5401
stars from them (18\%). Thus, $r_{ph}$ is more accurate
than $r_{HIP}$ and the results using $r_{ph}$ are the main ones
for 82\% of the sample stars.

\section*{RESULTS}

In Fig. 3, $R_V$ is plotted against the distance (with
a 100-pc step) for various latitude zones. The results
are indicated by the solid and dashed black lines for
the RGB stars with $r_{ph}$ and $r_{HIP}$, respectively, and by
the solid and dashed gray lines for the OB stars with
$r_{ph}$ and $r_{HIP}$, respectively. The choice of the latitude
zones is explained by the fact that the Gould Belt
passes in the zone $10^{\circ}<|b|<25^{\circ}$ and its influence
can manifest itself in the zone $25^{\circ}<|b|<40^{\circ}$, as was
shown previously (Gontcharov 2009).

In contrast to the OB stars, the RGB stars at high
latitudes ($|b|>40^{\circ}$) give fairly reliable results: the
number of RGB stars here is great and the mean
reddening $\overline{E_{(B_T-V_T)}}=0.1^m$ is sufficient for the
method to be applied.

The results for both classes of stars and both types
of distances agree within the errors limits indicated
by the vertical bars. This is an important result that
shows that, despite significant random errors, the
three independent distance scales (Hipparcos, RGB
photometry, and OB photometry) are in good agreement
in systematic terms. Below, we consider only
the results with $r_{ph}$.

Systematic $R_V$ variations between 2.7 and 4.0. are
seen at all latitudes. The minimum of $R_V$ is seen
on all plots at a distance from 250 to 550 pc. The
maximum of $R_V$ seen at a distance closer than 200 pc
from the Sun grows with increasing distance from the
Galactic equator. $R_V$ is almost constant along the
Galactic equator. All of this suggests the presence
of a vast, roughly symmetric (relative to the equator)
and nearly spherical (in shape) structure in the solar
neighborhood in which the dust grains are distributed
nonuniformly in their sizes and/or other properties
affecting $R_V$.

These systematic variations also manifest themselves
when analyzing the data in rectangular $XYZ$
coordinates. In this case, $R_V$ was calculated from
Eq. (1) for cells in the shape of a rectangular parallelepiped
with height $Z=100$ pc and a square base
the length of whose side (along $X$ or $Y$) was from
100 to 200 pc, so that there were at least 25 OB
stars and at least 36 RGB stars in the cell. The
calculations were performed by the moving average
method: in each step, the parallelepiped was shifted
by 20 pc along $X$ or $Y$ and $R_V$ was calculated for
the stars in the parallelepiped by the least-squares
method using Eq. (1). As a result, the calculations
were performed in more than 200 000 cells. Although
using a moving average slightly smoothed the results,
it is quite justified, because the distance errors are a
much more stronger smoothing factor (and the most
important source of systematic errors).

Figure 4 shows the contour maps of $R_V$ as a
function of the $X$ and $Y$ coordinates for three layers
along $Z$ for the RGB stars in the left column, for
the OB stars in the middle column, and averaged
for the two classes of stars in the right column:
(a) RGB, $+50<Z<+150$ pc;
(b) OB, $+50<Z<+150$ pc;
(c) mean $R_V$, $+50<Z<+150$ pc;
(d) RGB, $-50<Z<+50$ pc;
(e) OB, $-50<Z<+50$ pc;
(f) mean $R_V$, $-50<Z<+50$ pc;
(g) RGB, $-150<Z<-50$ pc;
(h) OB, $-150<Z<-50$ pc;
(i) mean $R_V$, $-150<Z<-50$ pc.
The gray scale for $R_V$ is given on the left. The isoline
step is $\Delta R_V=0.2^m$. The white lines of the coordinate
grid are plotted with a 500-pc step; the distance scale
on all plots is the same. The Sun is at the centers of
the plots; the Galactic center is on the right.

We see that the results for the two classes of stars
agree well, although fewer details are seen in the $R_V$
variations for the OB stars, because we had to use
large spatial cells (but with a side of no more than
200 pc, while the cell for the RGB stars is typically
$100\times100\times100$ pc). The standard deviation of the
``RGB minus OB'' $R_V$ differences within 500 pc of the
Sun is 0.2. This value may be considered an estimate
of the accuracy of the method that was achieved using
the data available to date. Obviously, the accuracy of
the result can be increased significantly by increasing
the accuracy of the photometry used and the distances.
The achieved accuracy of determining $R_V$
(0.2) in the presence of $R_V$ variations at least between
2.7 and 4 leaves no doubt that these variations are
real.

Below, we consider only the mean of the results for
the RGB and OB stars.

In the equatorial layer ($-50<Z<+50$ pc),
coarse dust dominates along the Great Tunnel, a
structure formed by clouds and young stars (Gontcharov
2004; Gontcharov and Vityazev 2005) -- the
light-gray band extending on plot (f) from the top
through the center to the lower left corner. The
region of coarse dust, along with the Gould Belt,
is inclined relative to the Galactic equator (by 17$^{\circ}$;
see Gontcharov 2009) and is located slightly northward of the equator toward the Galactic center (the
Scorpius--Sagittarius region is seen as a light spot at
the center of plot (c) in the first and fourth quadrants)
and southward toward the anticenter (the Perseus--Orion region is seen as light spots on plot (i) in the
second and third quadrants). The off-center position
of the Sun relative to the Gould Belt and the region of
coarse dust is also seen in the figure: the Scorpius--Sagittarius region with $r_{ph}\approx100$ pc and the Orion
region with $r_{ph}\approx400$ pc.

In fact, the main feature of the $R_V$ variations within
500 pc of the Sun is the radial, relative to the Sun
(to be more precise, apparently relative to the Local
Bubble and the center of the Gould Belt), gradient
in $R_V$ and, consequently, grain sizes. As we see
from the figure, $R_V$ is minimal in the Local Bubble,
a region of reduced gas density within about 100 pc
of the Sun. As the heliocentric distance increases, $R_V$
rises sharply and reaches its maximum immediately
outside the Local Bubble, at $r_{ph}\approx100\div150$ pc, in
complete agreement with the result by Skorzynski
et al. (2003), who detected a maximum of nonselective
extinction here. Still farther from the center of
this spherical structure, $R_V$ slowly decreases, reaching
its minimum, in fact, at the edge of the Gould Belt,
i.e., as has been shown above, at different heliocentric
distances, for example, in the direction of Scorpius--Sagittarius and Perseus--Orion. Outside the Gould
Belt, farther than 500 pc from the Sun, $R_V$ returns to
its mean value.

Apart from the variations associated with the
Gould Belt, a gradual increase in $R_V$ toward the
Galactic center, particularly in the equatorial layer, is
noticeable (plot (f)). In Fig. 5, the diamonds and the
solid line $R_V=0.3X+2.96$ indicate the dependence
of $R_V$ of the $X$ coordinate (in kpc) when averaging the
results in the layer $-50<Z<+50$, $-500<Y<+500$ pc. The dependence derived by Zasowski
et al. (2009) is also indicated here by the circles
and the dashed line. The vertical shift of the result
by Zasowski et al. (2009) can be caused by the
inaccuracy of the absolute calibrations of the infrared
extinction law mentioned above. Good agreement
between the revealed trends is important: 0.30 kpc$^{-1}$
in our study and 0.24 kpc$^{-1}$ in Zasowski et al.
(2009). The trend found is not an artefact, because
it does not manifest itself along $Y$ and $Z$. Thus, the
conclusion reached by Zasowski et al. (2009) about
the predominance of coarser dust at the Galactic
center is confirmed.

Figure 6 shows the contour maps of $R_V$ (the mean
of the results for the RGB and OB stars) as a function
of the coordinates: (a) $X$ and $Z$ for the layer
$-150<Y<+150$ pc; (b) $Y$ and $Z$ for the layer
$-150<X<+150$ pc. The gray scale for $R_V$ is
given on the right. The white lines of the coordinate
grid are plotted with a 500-pc step. the Sun is at the
centers of the plots.

We see that the central (roughly equatorial) layer
with a moderate $R_V\approx3.2\div3.4$ has a thickness corresponding
to the traditional thickness estimates for
the absorbing layer in the solar neighborhood -- about
150 pc (Gontcharov 2009). This means that if $R_V$ is
determined by the dust grain size, then it can be said
that fine and coarse dust dominates where there is
much and little dust, respectively. As in Figs. 3 and 4,
systematic $R_V$ variations in the Gould Belt and its
neighborhood manifest themselves here. In the $XZ$
plane on plot (a), the central layer with a moderate
$R_V\approx3.2\div3.4$ is inclined to the Galactic equator just
as the Gould Belt (the presence of this layer also
proves that the detected spherical structure in the $R_V$
variations is not the artefact that resulted from the
systematic errors in $r_{ph}$). The decrease in $R_V$ at $X\approx-600$ and $X\approx+250$ pc corresponds to the location
of the Gould Belt edges. Two regions of extremely
high $R_V$ clearly manifest themselves in this figure: two
roughly hemispheres northward and southward of the
Sun, with the northern one having a ``defect'' in the
second quadrant -- the region of lower $R_V$ independently
pointed out by Fitzpatrick and Massa (2007).
As the combination of Figs. 4 and 6 shows, there is
the previously detected, also roughly spherical layer
of minimum $R_V$ that also corresponds to the minima
in Fig. 3 at the outer boundary of these hemispheres
in all directions at a distance of about 500 pc from the
Sun. Still farther from the Sun, $R_V$ returns to some
mean value.

In general, the roughly radial (relative to the Sun)
variations of the extinction law found can be caused
by a systematic change in the composition of the
samples of stars with heliocentric distance. However,
the constancy of the normal $(B_T-V_T)_0$ color
for stars depending on the latitude, longitude, and
distance that we verified proves the invariability of the
composition of the samples.

It should be noted that there is no correlation, at
least pronounced one, between $R_V$ and the reddening
of stars: according to Gontcharov (2010), the reddening
in the region under consideration is maximal
in the first and second quadrants and has a deep
minimum in the third one, while $R_V$ is maximal in the
first and fourth quadrants and has a deep minimum in
the second one. As has been noted above, this was
detected long ago by Jonhson and Borgman (1963).
The absence of a correlation between the reddening
and $R_V$ also manifests itself in the fact that although
the Gould Belt contains dense clouds causing the
stars to redden (Gontcharov 2010), no $R_V$ variations
have been in the direction of these clouds that would
differ significantly from those in other directions.

\section*{CONCLUSIONS}

This is the third paper in our series of studies of
the interstellar extinction in the Galaxy. Our previous
studies (Gontcharov 2009, 2010) showed that the
stellar reddening and extinction could be analyzed using
accurate multicolor broadband photometry from
present-day surveys of millions of stars. In addition,
these studies revealed a great role of the Gould Belt as
a region containing absorbing matter and, in addition,
oriented almost radially relative to the Sun.

Our investigation has shown that the hard-to-implement
extinction law extrapolation method proposed
about 50 years ago can be implemented at
present in both infrared and visual spectral ranges,
for example, using multicolor broadband photometry
from the Tycho-2 and 2MASS catalogues for 11 990
OB and 30 671 K-type RGB stars and photometric
distances. The values of $R_V$ obtained agree for the
two classes of stars so that the standard deviation of
the differences is 0.2. This accuracy level allows consistent
(for the two classes of stars) systematic largescale
variations of the extinction law and, accordingly,
the ratio $R_V$ and the mean dust grain size within
the nearest kiloparsec to be revealed with confidence.
The detected $R_V$ variations between 2.2 and 4.4 not
only manifest themselves in individual clouds but also
span the entire space near the Sun, being related to
the main Galactic structures of the nearest kiloparsec.
In the Local Bubble within about 100 pc of
the Sun, $R_V$ has a minimum. In the inner part of
the Gould Belt and at high Galactic latitudes, at a
distance of about 150 pc from the Sun, $R_V$ reaches
a maximum and then decreases to its minimum in the
outer part of the Belt and other directions at a distance
of about 500 pc from the Sun, returning to its mean
values far from the Sun.

Near the Galactic equator, we found a monotonic
increase in $R_V$ by 0.3 per kpc toward the Galactic
center that is consistent with the result obtained by
Zasowski et al. (2009) for much of the Galaxy.

We showed that ignoring the $R_V$ variations and
traditionally using a single value for the entire space
should lead to systematic errors in the calculated
distances reaching 10\%. In addition, the detection
of such large $R_V$ variations forces us to recalculate
the data of the numerous reddening and extinction
maps obtained in recent years from the photometry of
millions of stars.

The dependences of $R_V$ on the reddening, extinction,
spectral type, and other characteristics of stars
found by various researchers (Straizys 1977) are not
fully explained by the theory of the method under
consideration. The consistency of the $R_V$ variations
for red giants and OB stars found here suggests that
many of these dependences, if not all, are the artifacts
that resulted from the disregarded correlations
between the stellar characteristics and $R_V$ variations.

In this study, we independently confirm the radial
(relative to the Sun) orientation of the absorbing
matter within the nearest kiloparsec found previously
(Gontcharov 2010). The Local Bubble and the central
region of the Gould Belt located near the Sun seem to
be actually the center here. This, along with the detection
of regions with extremely large $R_V$ and, hence,
large nonselective extinction at high latitudes, is of
great importance for estimating the extinction toward
extragalactic objects and the mass of the baryonic
dark matter.

This paper may be considered as groundwork at
the beginning of extensive studies, primarily as a
proof of the necessity of investigating the large-scale
variations of the extinction law. Analysis of the difficulties
in applying the extinction law extrapolation
method revealed not only its great potential but also
the need for original data of a completely different
level. At present, there is no accurate infrared photometry
for bright stars and visual all-sky photometry
for stars fainter than $11^m$. Only the largest-scale
variations of the extinction law and only within about
500 pc of the Sun can be revealed by applying the
method to tens of thousands of stars. At the same
time, our investigation showed that the influence of
the Gould Belt and other local structures disappears
only beyond this radius and precisely the Galactic
variations of the extinction law can at last manifest
themselves. Complete samples of stars within several
kpc of the Sun (i.e., samples of tens of millions of
stars to $18^m$) are needed for their serious analysis.

The number of stars is particularly small at high
latitudes, precisely where the most interesting results,
especially important for extragalactic astronomy and
dark matter studies, are evident. As a result, the mean
accuracy of the photometry in a spatial cell at high
latitudes barely exceeds the reddening. This does not
allow the final conclusions to be reached. Since the
layer of OB stars is thin, the extinction law at high
latitudes can apparently be investigated only by using
red giants.

Thus, not only the parallaxes from the Gaia project
but also the reconstruction of the entire complex
wavelength dependence of extinction in each spatial
cell are needed to construct an accurate three-dimensional
map for the variations of the extinction
law in much of the Galaxy. In turn, this requires
multiband photometry (in fact, spectrophotometry)
with an accuracy of at least 0.01$^m$ in the ultraviolet,
visual, and infrared ranges for tens of millions of red
giants with $0^m<V<18^m$ over the entire sky.

\section*{ACKNOWLEDGMENTS}

In this study, we used results from the Hipparcos and 2MASS (Two Micron All Sky Survey)
projects, the SIMBAD database (http://simbad.ustrasbg.fr/simbad/), and other resources of the Strasbourg
Data Center (France), http://cds.u-strasbg.fr/. The study was supported by the ``Origin and Evolution
of Stars and Galaxies'' Program of the Presidium of the Russian Academy of Sciences.

\begin{figure}[h]
\includegraphics{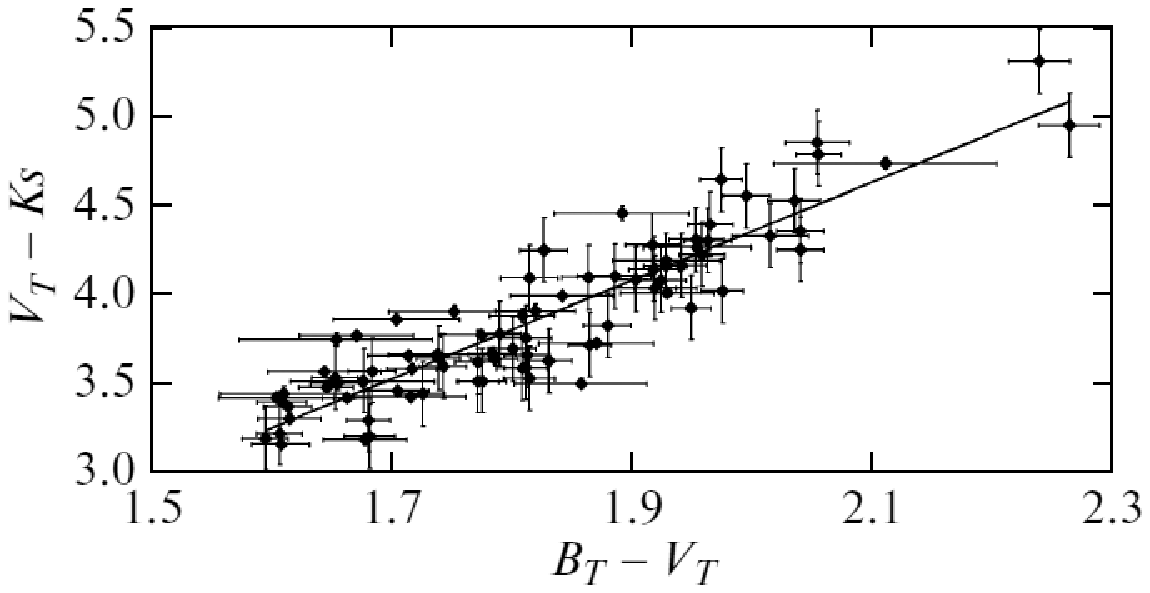}
\caption{$(V_T-Ks)$ -- $(B_T-V_T)$ diagram for K-type RGB
stars in the spatial cell with $0^{\circ}<l<90^{\circ}$, $-10^{\circ}<b<+10^{\circ}$, $200<r_{HIP}<300$ pc.
}
\label{bvvk}
\end{figure}

\begin{figure}[h]
\includegraphics{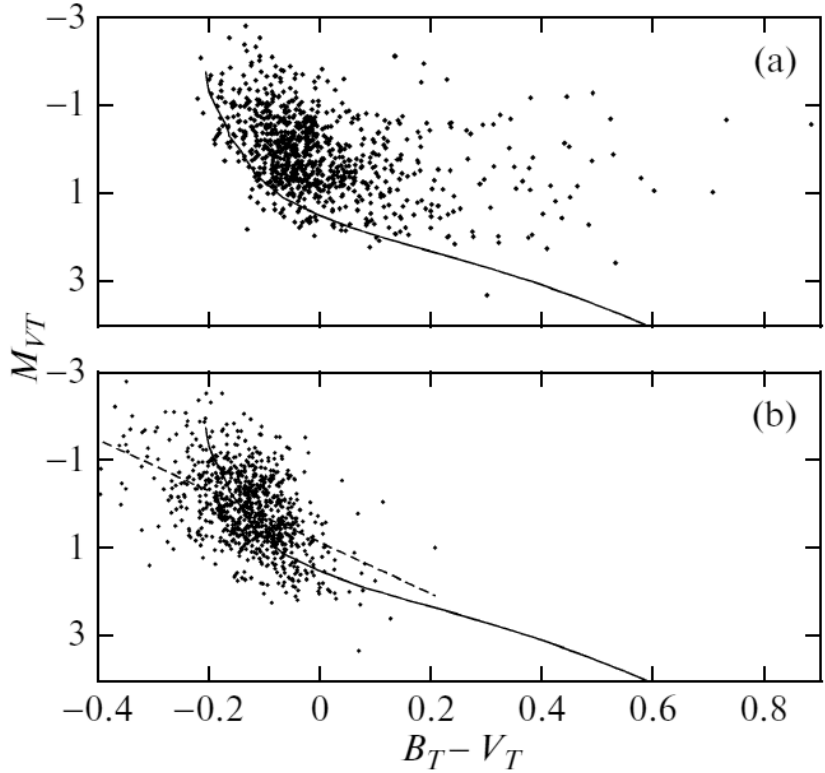}
\caption{Distribution of 871 OB stars within 300 pc of
the Sun on the $(B_T-V_T)$ -- $M_{V_T}$ diagram: (a) with the
original $(B_T-V_T)$, (b) with the dereddened $(B_T-V_T)_0$.
The solid and dashed curves represent the isochrone for
solar-metallicity stars with an age of 60 Myr and the
adopted calibration, respectively.
}
\label{btvtmvt}
\end{figure}

\begin{figure}[h]
\includegraphics{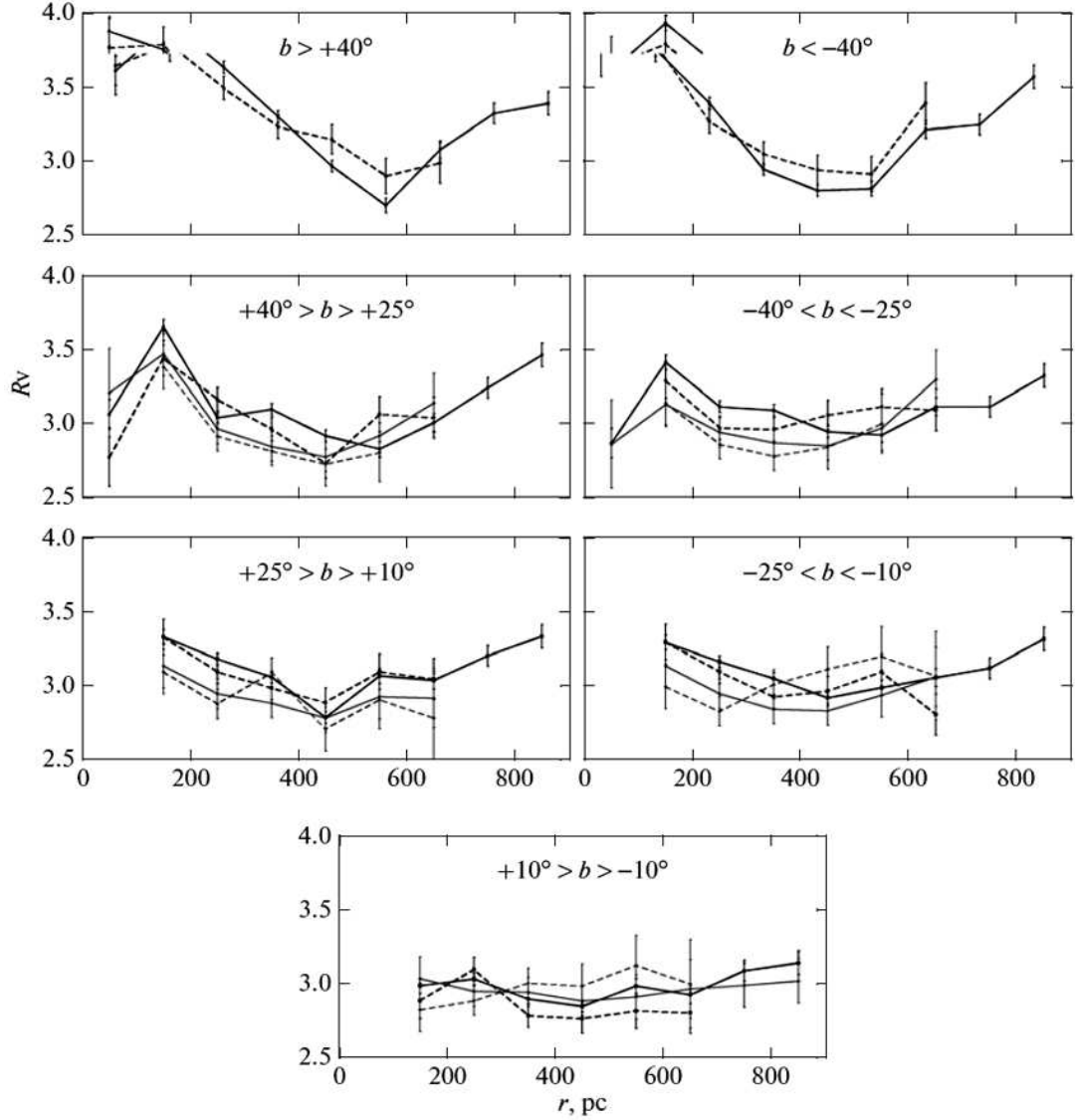}
\caption{$R_V$ versus distance for various Galactic latitudes: the photometric and
trigonometric distances of the RGB stars are
indicated by the solid and dashed black lines, respectively; the photometric and
trigonometric distances of the OB stars are
indicated by the solid and dashed gray lines, respectively.
}
\label{high}
\end{figure}

\begin{figure}[h]
\includegraphics{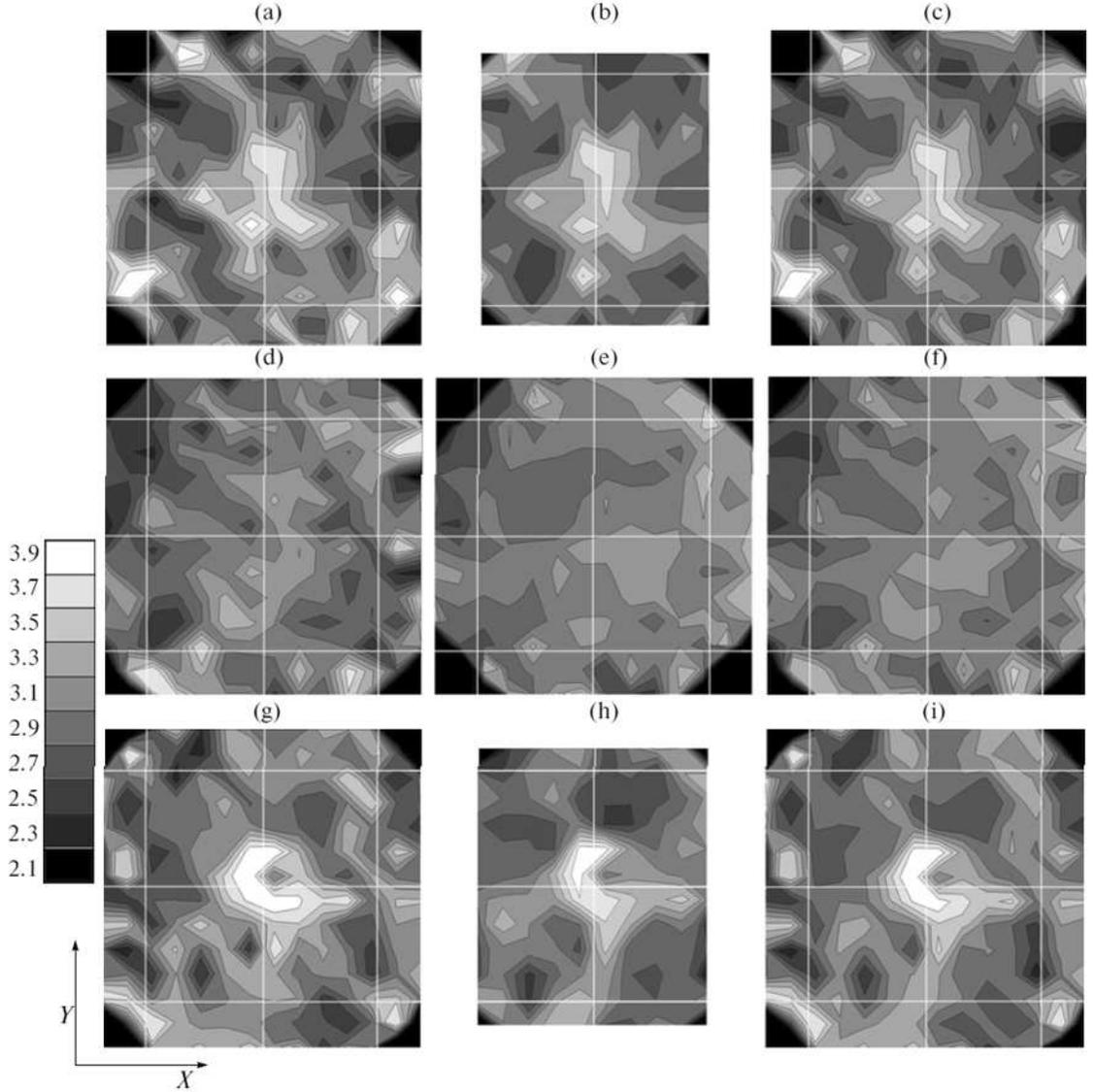}
\caption{Contour maps of $R_V$ as a function of the $X$ and $Y$ coordinates:
(a) RGB, $+50<Z<+150$ pc;
(b) OB, $+50<Z<+150$ pc;
(c) mean $R_V$, $+50<Z<+150$ pc;
(d) RGB, $-50<Z<+50$ pc;
(e) OB, $-50<Z<+50$ pc;
(f) mean $R_V$, $-50<Z<+50$ pc;
(g) RGB, $-150<Z<-50$ pc;
(h) OB, $-150<Z<-50$ pc;
(i) mean $R_V$, $-150<Z<-50$ pc.
The gray scale for $R_V$ is shown on the left. The white lines of the coordinate grid are plotted with a
500-pc step. The Sun is at the centers of the plots; the Galactic center is on the right.
}
\label{cube}
\end{figure}

\begin{figure}[h]
\includegraphics{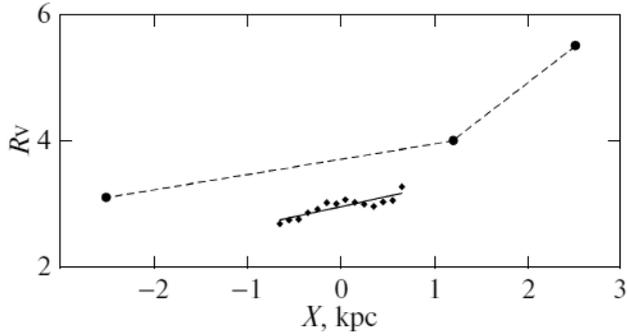}
\caption{Dependence of $R_V$ on $X$ coordinate: obtained
here when averaging the results in the layer $-50<Z<+50$ pc, $-500<Y<+500$ pc (the diamonds
and solid line); obtained by Zasowski et al. (2009) (the
circles and dashed line).
}
\label{rrv}
\end{figure}

\begin{figure}[h]
\includegraphics{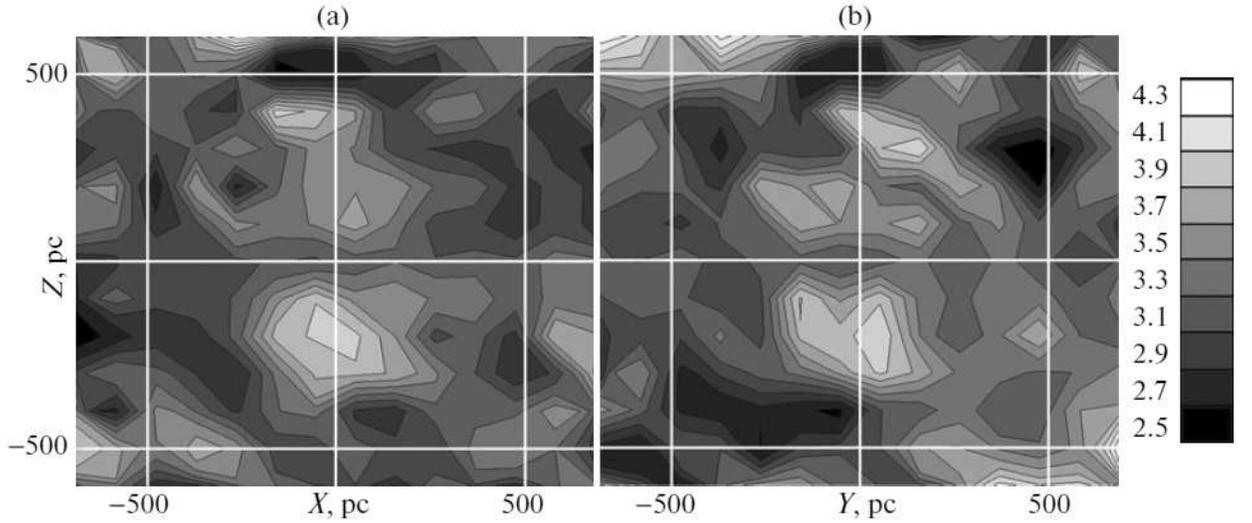}
\caption{Contour maps of $R_V$ as a function of the coordinates: (a) $X$ and $Z$ for the layer $-150<Y<+150$ pc;
(b) $Y$ and $Z$ for the layer $-150<X<+150$ pc.
The gray scale for $R_V$ is given on the right. The white lines of the coordinate grid are
plotted with a 500-pc step; the Sun is at the centers of the plots.
}
\label{xyz}
\end{figure}

\end{document}